# Pronounced Site Preference in Cr-Doped Mn-Based M-Type Hexaferrites and Its Chemical Origins

*Dylan Correll,[a] Susheng Tan,[b] Evan Wang[a] and Xin Gui[a]**

[a] Department of Chemistry, University of Pittsburgh, Pittsburgh, PA, 15260, USA
[b] Department of Electrical and Computer Engineering and Gertrude E. and John M. Petersen Institute of Nanoscience and Engineering, University of Pittsburgh, Pittsburgh, PA, 15260, USA
Address correspondence to E-mail: xig75@pitt.edu

## *Abstract*

Cation distribution plays a critical role in determining the properties of crystalline oxides. Understanding the chemical factors that govern cation distribution is therefore essential for the rational design of cation-ordered materials, particularly frustrated magnets in which structural disorder can strongly affect the magnetic ground state. Here, we investigate the origin and evolution of cation distribution in the extensively studied M-type hexaferrite structure using the first Mn-based M-type hexaferrite, $KSb_3Mn_9O_{19}$, as a model system. A systematic $Cr^{3+}$-doped series, $KSb_3(Mn_{1-x}Cr_x)_9O_{19}$, was examined using powder and single-crystal X-ray diffraction, electron microscopy, and energy-dispersive X-ray spectroscopy. $Cr^{3+}$ exhibits a pronounced preference for the Mn Kagome sublattice, accompanied by the evolution of Mn vacancies and $Mn^{3+}/Sb^{3+}$ disorder on neighboring sites. Analysis of the local coordination environments and their evolution with Cr content suggests that crystal-field effects, chemical bonding, and local structural strain collectively govern this site selectivity. The magnetic properties of the Cr-doped compounds show similar behaviors as the undoped parent compound, other than the doping-induced spin-glass state at the highest Cr concentration, supported by heat capacity measurements. These results establish a chemical picture of cation site distribution and selection in M-type hexaferrites and demonstrate how substitution at one crystallographic site can induce coupled redistribution and disorder across neighboring sublattices. Moreover, this work establishes local coordination chemistry as a route toward understanding and ultimately controlling cation distribution in complex oxides, providing chemical design principles for structurally well-defined frustrated magnetic materials.

## *Introduction*

Frustrated magnetism arises in materials where competing magnetic interactions cannot be simultaneously satisfied, resulting in the absence of long-range magnetic order relative to the energy scale of the magnetic interactions.[1–5] A common origin of magnetic frustration is geometrical frustration, where magnetic species occupy specific geometries, including triangular,[6–10] Kagome[11–16] and related[17–20] lattices. When combined with strong quantum fluctuations, geometrical frustration can stabilize unconventional magnetic ground states such as quantum spin liquids and quantum spin ices.[21–25] Experimentally identifying such states, however, remains challenging because cation disorder can introduce random local environments and exchange interactions that obscure the intrinsic response of the frustrated magnetic lattice.[26–28] For example, cation disorders in $YbMgGaO_4$ and $ZnCu_3(OH)_6Cl_2$, i.e., $Mg^{2+}/Ga^{3+}$ and $Cu^{2+}/Zn^{2+}$ disorders,[29–33] complicate the interpretation of their magnetic ground states using current techniques, making it difficult to definitively determine the existence of a quantum frustrated magnetism state. Additionally, cation disorders can also lead to spin-glass states that further avoid the determination of quantum frustrated magnetism e.g., $SrCr_{9p}Ga_{12-9p}O_{19}$[34–37]. Therefore, understanding how such cation distributions and disorders are formed is critical for the rational design of structurally well-defined frustrated magnets and for distinguishing intrinsic frustration-driven behavior from disorder-induced phenomena.

M-type hexaferrites with the general formula of $AB_{12}O_{19}$ provide an attractive platform for investigating cation distribution in structurally complicated oxides. They adopt five crystallographically distinct B sublattices, including geometrically frustrated triangular and Kagome sublattices.[38–40] Although extensive chemical substitution is possible on the B sites, different B-site cations commonly distribute over multiple crystallographic sites, and M-type hexaferrites with complete or nearly complete B-site cation order remain uncommon.[41–43] For example, substituted cations such as Co and Al have been found to occupy several of the five B sites despite showing distinct site preferences. Among over 150 reported M-type hexaferrite phases in the ICSD, only a few were reported to host B-site frustrated magnetism; however, they were all spin glass materials due to the extensive B-site cation disorders, e.g., $M(Cr_{9p}Ga_{12-9p})O_{19}$ (M = Sr, Ba),[44–49] $Sr(Co_6Ti_6)O_{19}$[50,51] and $Ba(Co_6Sn_6)O_{19}$,[52] making them chemically challenging to control, characterize and interpret cation distribution and frustrated magnetism. In contrast, the recently reported first Mn-based M-type hexaferrites, $ASb_3Mn_9O_{19}$ (A = K or Rb),[40] significantly suppressed cation disorders, exhibits distinct Mn and Sb sublattices and therefore, provides a useful starting point for understanding how B-site cation order can be established and maintained in the M-type hexaferrites. Here, $Mn^{3+}$ ions occupy three sublattices with distinct coordination environments, i.e., Mn1 triangular (trigonal bipyramidal), Mn2 puckered honeycomb (tetrahedral) and Mn3 Kagome (octahedral). Moreover, $ASb_3Mn_9O_{19}$ was found to host quasi-two-dimensional frustrated magnetism, representing an opportunity to systematically investigate both the chemical factors governing cation distribution in complex structures, and the relationship between cation distribution and frustrated magnetism.

In this work, we performed systematic $Cr^{3+}$ doping on $KSb_3Mn_9O_{19}$ and conducted comprehensive structural, magnetic and thermodynamic characterizations. Structural refinements and atomic-resolution STEM-EDS suggest pronounced site preference of $Cr^{3+}$ onto the Mn3 Kagome site, while $Mn^{3+}$/$Sb^{3+}$ mixing and $Mn^{3+}$ vacancies gradually evolve on the Mn1 and Mn2 sites, respectively. The strong site preference is particularly notable, given the identical formal charge and similar ionic radii of $Cr^{3+}$ and $Mn^{3+}$. Analysis of the local coordination environments, bond lengths, bond angles, and structural distortions suggests that crystal-field effects, chemical bonding, and local strain collectively determine the $Cr^{3+}$ site preference. These results show that cation distribution in M-type hexaferrites cannot be explained by ionic size alone and that substitution on one site can also induce cation disorder and vacancies on the surrounding sites. Understanding these chemical factors provides guidance for directing different cations toward specific B-site sublattices and may ultimately enable the design of new B-site-ordered M-type hexaferrites. Such control is particularly important for frustrated magnetic systems, where cation disorder can strongly affect the magnetic ground state.

## *Experimental Details*

### Synthesis of polycrystalline $KSb_3(Mn_{1-x}Cr_x)_9O_{19}$

Polycrystalline $KSb_3(Mn_{1-x}Cr_x)_9O_{19}$ (x = 3.33%, 6.67%, 10%, 13.3%, 16.7%, and 25%) was prepared using a high temperature solid-state method. A stoichiometric mixture of $Sb_2O_3$ (99.9%, Thermo Scientific), MnO (99.99%, Thermo Scientific) and $Cr_2O_3$ (99.6%, Thermo Scientific) with 60% excess anhydrous $K_2CO_3$ (99+%, Thermo Scientific), were thoroughly ground and placed in alumina crucibles. An excess of $K_2CO_3$ was used to compensate for its vaporization under high temperatures.

The mixture was heated to 1200 °C and kept for 12 hours. All samples were re-annealed at the same temperature overnight several times with intermittent grindings until phase pure products were obtained. The samples were then washed with water to remove any excess $K_2CO_3$.

**Single Crystal and Powder XRD**

Single crystal X-ray diffraction (SCXRD) studies were carried out on a Bruker D8 Quest ECO diffractometer equipped with APEX5 software and Mo radiation ($\lambda_{K\alpha}$ = 0.71073 Å). The crystals were soaked in glycerol and mounted on a Kapton loop. The direct method and full-matrix least-squares on F2 procedure within the SHELXTL package were employed to solve the crystal structure.[53,54]

Powder X-ray diffraction (PXRD) measurements were utilized to determine the phase purity for the polycrystalline samples. A Bruker D2 PHASER with Cu Kα radiation (λ = 1.54060 Å, Ge monochromator) and a LynxEye-XE detector was employed. The Bragg angle measured was from 5 to 100° at a rate of 1.7°/min with a step of 0.012°.

**Physical Properties Measurement**

The Quantum Design Dynacool Physical Property (PPMS) was used to measure the DC and AC magnetization from 2 to 300 K under various applied external magnetic fields using the equipped ACMS II option. Field-dependent magnetization data were collected in the range −9 to 9 T under different temperatures. Temperature dependence of magnetic susceptibility was measured from 2 to 300 K under an external magnetic field of 0.3 T. Zero field cooling (ZFC) protocols were applied. Heat capacity was measured using a standard relaxation method in the PPMS from 1.9 to 100 K. Magnetic properties were performed on the water washed powder samples. Heat capacity measurements were conducted using pelletized samples.

**Transmission Electron Microscopy (TEM)**

Samples were prepared using Thermo Scientific SCIOS focused ion beam and scanning electron microscope dual beam system equipped with field-emission electron gun, Ga ion source, Carbon and Platinum gas injectors, and in-situ lift-out manipulator. Transmission electron microscopy (TEM) characterization was carried out with a Thermo Scientific Titan Themis G2 200 probe aberration-corrected TEM equipped with SuperX EDS spectrometer and operated at 200 kV in both TEM and STEM modes.

**X-ray Photoelectron Spectrometry (XPS)**

X-ray Photoelectron Spectrometry (XPS) was performed on select samples using an ULVAC-PHI (Physical Electronics) GENESIS XPS with Al Kα radiation (1486 eV). The pass energy was 200 eV for the wide (survey) spectra and 50 eV for the high-resolution regions (narrow spectra). The base pressure of the analysis chamber was less than $\approx 1 \times 10^{-9}$ mbar. The analysis chamber pressure was at $1 \times 10^{-7}$ mbar during data acquisition. Polycrystalline sample was brushed on to carbon tape for XPS measurements.

**Scanning Electron Microscopy (SEM)-Energy-Dispersive X-ray Spectroscopy (EDS)**

Compositional analysis was performed via scanning electron microscopy (SEM) with energy-dispersive spectroscopy (EDS) on a Zeiss Sigma 500 VP SEM instrument with Oxford Aztec X-EDS. Samples were mounted on carbon tape before loading into the SEM chamber. Multiple points and areas were examined from each sample to get the K:Sb:Mn:Cr ratio. Samples were analyzed at 20 kV.

## *Results and Discussion*

**Phase purity and crystal structure of $KSb_3(Mn_{1-x}Cr_x)_9O_{19}$:** The phase purity of $KSb_3(Mn_{1-x}Cr_x)_9O_{19}$ (x = 0%, 3.33%, 6.67%, 10%, 13.3%, 16.7% and 25%) is confirmed using powder X-ray diffraction (XRD), as shown in Figure 1a. No obvious additional peaks can be seen in the powder XRD patterns, while the Cr-doped compounds are consistent with undoped $KSb_3Mn_9O_{19}$, indicating the correct phases crystallizing in the space group $P6_3/mmc$, and high phase purity throughout the series. Systematic peak shifts are observed, as evidenced by the inset of Figure 1a demonstrating the evolution of the 220 peak. Le Bail fitting[55] is employed to determine the lattice parameters from powder XRD patterns. As can be seen in Figure 1b, the lattice parameter $a$ is found to decrease with increasing $x$, consistent with the smaller cationic size of $Cr^{3+}$ compared to $Mn^{3+}$. However, the lattice parameter $c$ exhibits a slight upturn at low $x$ followed by a monotonic decreasing trend with high $x$. The high phase purity and systematic $a/c$ trend suggest that $Cr^{3+}$ has been successfully doped into $KSb_3Mn_9O_{19}$. The fact that the starting material, $Cr_2O_3$, possesses high melting and boiling points further suggests the consistency of Cr contents in the final products and the loading compositions.

The crystal structures of $KSb_3(Mn_{1-x}Cr_x)_9O_{19}$ are determined using single-crystal XRD at room temperature. They all appear to be consistent with the reported M-type hexaferrites structure.[40] The crystallographic data including atomic sites, site occupancies, refined anisotropic displacement parameters and equivalent isotropic thermal displacement parameters of all compounds are summarized in Tables S1 – S3 in the supporting information (SI). The overall crystal structure is shown in Figure 1c, containing three Mn sites and two Sb sites. The three Mn sites form triangular (Mn1), puckered honeycomb (Mn2) and Kagome (Mn3) sublattices, while they possess five- (trigonal bipyramidal), six- (octahedral), and four-(tetrahedral) coordination, respectively. The two Sb sites are both six-coordinated in $Sb@O_6$ octahedra. Since Cr and Mn are closely placed in the periodic table, it is challenging for the lab-based single-crystal X-ray diffractometer to resolve the two cations. Therefore, all Mn and Sb sites were initially relaxed in our refinement models. By doing so, Sb1 and Mn1 sites were found fully occupied within the standard deviations, while consistent lower-than-unity occupancies can be seen on Mn2, Mn3 and Sb2 sites, as shown in Table S4 in the SI, where the vacancies are from 7.1% to 14.3% on the Mn2 site, 0.5% to 3.2% on the Mn3 site and 7.3% to 26.7% on the Sb2 site. Therefore, the major focus on the structural analysis will be on Mn2, Mn3 and Sb2 sites, as demonstrated below.

The possible origins of the refined vacancies for the Mn2 and Mn3 sites can be twofold: missing $Mn^{3+}$ ions, or the substitution of $Mn^{3+}$ by $Cr^{3+}$ ions. For Sb2 site, missing $Sb^{3+}$ ions or substitution of $Sb^{3+}$ by $Cr^{3+}/Mn^{3+}$ ions may potentially be attributed to the lower-than-unity occupancy. Given the $Mn^{3+}/Sb^{3+}$ disordered nature on the Sb2 site in the parent compound, $KSb_3Mn_9O_{19}$,[40] the substitution with $Cr^{3+}/Mn^{3+}$ ions is likely to be the major reason. However, single-crystal XRD alone cannot provide sufficient evidence for accurate crystal structures. Nevertheless, the similar trend of lattice parameters between powder and single-crystal XRD results shown in Figure 1b validates the XRD refinement results.

**Site-selective doping of $Cr^{3+}$, evolving Mn2-site vacancies and Mn/Sb disorder:** In order to obtain more in-depth information about the crystal structures and site occupancies of $KSb_3(Mn_{1-x}Cr_x)_9O_{19}$, atomic-resolution STEM-EDS was performed on nominal $KSb_3(Mn_{0.9}Cr_{0.1})_9O_{19}$, i.e., x = 10%. As shown in Figure 2a, Figure 2d, and Figure S1 in the SI, two different regions of a $KSb_3(Mn_{0.9}Cr_{0.1})_9O_{19}$ crystal were tested for Energy-dispersive X-ray spectroscopy (EDS) line profiles and High-Angle Annular Dark-Field (HAADF) imaging, respectively. The HAADF images (zone axis $[11\bar{2}0]$) in Figures 2d and S1 clearly confirm the refined unit cell using single-crystal XRD, while the EDS mapping suggests even distribution of K and Sb elements. However, they exhibit relatively broadened elemental contrast for Mn and Cr, likely arising from the intrinsically limited X-ray counting statistics of atomic-scale EDS, together with sample drifting during the data acquisition. Therefore, the analyses on the site occupancies will concentrate on the line profiles obtained from the STEM-EDS measurements.

*Pronounced site preference of $Cr^{3+}$ substitution on the Mn3 Kagome site:* Figure 2a shows the line profiles based on the HAADF images in Figures 2d and S1, where the vertical dashed lines indicate the positions of the Mn2 puckered honeycomb and Mn3 Kagome sites. Clearly, the periodic double peaks of Mn3 Kagome sites can be found corresponding to the repeated double peaks of Cr at the same position for both areas. This suggests that $Cr^{3+}$ ions are consistently doped onto the Mn3 Kagome site, which is in good agreement with the slight vacancies on Mn3 site observed from single-crystal XRD data in Table S4 in the SI. Note that Cr (Z = 24) and Mn (Z = 25) are close in the periodic table while $Z_{Cr}/Z_{Mn}$ = 96%, this validates the small vacancies from the XRD refinements. The single-crystal XRD results also suggest much more significant vacancies on the Mn2 and Sb2 sites that are unlikely to be entirely due to the $Cr^{3+}$ doping. Instead, no obvious peaks of $Cr^{3+}$ can be observed in Figure 2a at the Mn2 puckered honeycomb positions marked by the purple dashed vertical lines in either area, indicating no/minor amount of $Cr^{3+}$ on the Mn2 site. Additionally, a peak in between the two Mn2 atoms represents the Mn occupancy on the Sb2 site. However, no clear peak can be seen for $Cr^{3+}$ at the corresponding position, suggesting zero/minor appearance of $Cr^{3+}$ on the Sb2 site. Although the STEM-EDS data cannot provide definitive conclusion that $Cr^{3+}$ solely occupies the Mn3 Kagome site due to the data noise mentioned above, it is plausible to suggest that much more significant $Cr^{3+}$ occupancy is observed on the Mn3 site compared to Mn2 and Sb2 sites, revealing pronounced site preference of $Cr^{3+}$ doping in $KSb_3Mn_9O_{19}$.

*Evolving site vacancy on the Mn2 puckered honeycomb site:* Single-crystal XRD refinements suggest clear vacancies on the Mn2 site, while the analysis of $Cr^{3+}$ occupancy above indicates zero or minor appearance on the Mn2 site, suggesting that the major contributor to the loss of XRD peak intensities is site vacancy, i.e., missing $Mn^{3+}$ ions. This is further proved by analyzing the atomic fraction of the Mn2 puckered honeycomb site in Figure 2a. There are eight groups of Mn2 sites tested, i.e., 1*a* – 1*d* in area 1 and 2*a* – 2*d* in area 2. Assuming a fully occupied Mn2 site, two equal peaks within the detection limit should be observed in all groups. However, only groups 1*b*, 1*d* and 2*c* exhibit similar atomic fraction, while other groups show "asymmetric" peak heights, suggesting different occupancies of $Mn^{3+}$ in these groups, which is consistent with $Mn^{3+}$ vacancies on the Mn2 site. With this, the occupancy from the single-crystal XRD refinement is plotted in the lower panel of Figure 2b with a monotonic, but nonlinear trend with increasing $x$.

*$Mn^{3+}/Sb^{3+}$ disorder on the Sb2 site:* In the parent compound, $KSb_3Mn_9O_{19}$, possible $Mn^{3+}/Sb^{3+}$ disorder was reported on the Sb2 site.[40] However, the conclusion was only based on single-crystal XRD and SEM-EDS results such that a definitive answer is missing. Here for $KSb_3(Mn_{0.9}Cr_{0.1})_9O_{19}$, Figure 2a consistently shows a peak for the atomic fraction of Sb in between two purple dashed lines for Mn2 site, suggesting the appearance of $Sb^{3+}$ on the Sb2 site. Moreover, a peak in the atomic fraction of Mn can be seen at the corresponding position in groups 1*a*, 1*c*, 1*d*, 2*c* and 2*d*, suggesting $Mn^{3+}/Sb^{3+}$ disorder on the Sb2 site. The Mn peak intensity is also found missing or suppressed in other groups, consistent with the disordered nature and the fact that $Mn^{3+}$ is the minor component on the Sb2 site. Based on the single-crystal XRD refinement, the occupancy of $Mn^{3+}$ on the Sb2 site is plotted in the upper panel of Figure 2b, revealing a non-monotonic trend with increasing $x$.

To summarize the crystal structure after analyzing single-crystal XRD and STEM-EDS results, Figure 2c presents a general crystal structure of $KSb_3(Mn_{1-x}Cr_x)_9O_{19}$, for which the chemical disorder mainly exists in the Mn3-Mn2-Mn3 tri-layer blocks. Additionally, the chemical formula can be rewritten as a site-specific version, i.e., $KSb_2(Sb_{1-y}Mn_y)Mn_1(Mn_{1-\delta})_2(Mn_{1-x}Cr_x)_6O_{19}$.

**Proposed mechanisms for site preference, vacancies and disorders:** Cationic size is one of the most widely studied chemical factors governing the degree of cation disorder in crystalline solids, e.g., in double perovskites,[56–58] pyrochlores,[59–61] and spinels.[62–64] However, it may not effectively explain the pronounced site preference of $Cr^{3+}$ in $KSb_3(Mn_{1-x}Cr_x)_9O_{19}$, due to the similar cationic sizes between $Cr^{3+}$ and $Mn^{3+}$, i.e., $r_{Cr^{3+}} \sim 0.615$ Å *vs* $r_{Mn^{3+}} \sim 0.645$ Å (octahedral; high-spin).[65] Moreover, the interpretation of the evolving vacancies on the Mn3 site and the varied $Mn^{3+}/Sb^{3+}$ mixing on the Sb2 site is still lacking. Therefore, other potential mechanisms need to be proposed to better understand the chemical order-disorder conditions in the extensively studied M-type hexaferrites family, and to assist with future exploration of the chemically ordered M-type hexaferrites.

*Crystal-field- and chemical-bond-assisted site preference of $Cr^{3+}$ on the Mn2 site:* Given the similar ionic size and electronegativity of $Cr^{3+}$ and $Mn^{3+}$, one may expect $Cr^{3+}$ to be evenly distributed on all three Mn sites and the Sb2 site, if ionic size is the only determining factor for site selectivity, which is apparently not true in $KSb_3(Mn_{1-x}Cr_x)_9O_{19}$. By comparing Mn3 Kagome site with Mn1 and Mn2 sites, one prominent difference is the coordination number of the three sites, i.e., Mn1 (5-coordinated trigonal bipyramidal), Mn2 (4-coordinated tetrahedral) and Mn3 (6-coordinated octahedral) as can be seen in Figure 1c. Therefore, crystal-field stabilization energy (CFSE) can be one of the leading factors determining whether $Cr^{3+}$ occupies these three sites. As shown in Figure 3a, the CFSE for substituting Mn1, Mn2 and Mn3 sites with $Cr^{3+}$ ($S = 3/2$) ion are ~ -0.63 $\Delta_O$, ~ -0.36 $\Delta_O$, and ~ -1.2 $\Delta_O$, respectively, suggesting that crystal-field effect can stabilize the $Cr^{3+}$ doping onto the octahedrally coordinated Mn3 Kagome site more significantly than the other two sites, consistent with our experimental observations.

However, CFSE may not be able to fully explain the site preference of $Cr^{3+}$ on the Mn3 site over the Sb2 site, which is also octahedrally coordinated and can be occupied by $Mn^{3+}$. Moreover, the Sb2 site is less distorted compared to the Mn3 site, which is favored by a $d^3$ ion. This indicates that additional factors other than CFSE are needed to explain the site preference. It can be found that the reported $Cr^{3+}$-$O^{2-}$ bond lengths in a Cr@$O_6$ octahedron mostly lie within the range of 1.98 – 2.09 Å, e.g., the Cr-based spinels.[66–68] Here, Table 1 shows important metal-oxygen bond lengths across the series of Cr-doped $KSb_3Mn_9O_{19}$, while Figure 3b presents the representative region of interest with atom labels. Surprisingly, although systematic lattice parameter shift is observed in Figure 1b, Sb2-O1 bond length remains nearly the same with $Cr^{3+}$ doping (from 1.977 Å when x = 0% to 1.973 Å when x = 25%), indicating that the Sb2 site is robust against local structural distortion/perturbation. In contrast, the Mn3-O bond lengths range from ~1.98 Å to ~ 2.17 Å, suggesting a higher flexibility in accommodating ions with different sizes. Given that the Sb2-O1 bond is slightly below or at the lower end of the previously reported $Cr^{3+}$-$O^{2-}$ bond lengths, it is plausible to believe that $Cr^{3+}$ ions preferentially occupy Mn3 site over Sb2 site due to the restriction on the chemical bond lengths in the Sb2@$O_6$ octahedron.

*Strain-induced Mn2-site vacancies:* The Mn2-site vacancy is likely correlated with $Cr^{3+}$ doping since no Mn2-site vacancy was observed in the parent compound, $KSb_3Mn_9O_{19}$.[40] Moreover, Mn2@$O_4$ tetrahedra are sandwiched by two layers of Mn3@$O_6$ octahedra. Thus, $Cr^{3+}$ doping onto the Mn3 site will highly likely affect the local structure and bonding of Mn2. Evolution of the Mn2-O bond lengths can be found in Table 1, while Table 2 summarizes the changes on the distorted $Mn2@O_4$ tetrahedron with increasing $Cr^{3+}$ doping. It can be found that Mn2-O1 exhibits a reduced bond length with increasing $x$, while Mn2-O4 bond lengths remains similar and drop at $x$ = 25%. Notably, solely based on the evolution of bond lengths, $Mn2@O_4$ shifts from a distorted tetrahedron to a nearly "ideal-tetrahedron" scenario, i.e., from $d_{Mn2-O1}/d_{Mn2-O4}$ ~ 97.9% at x = 0% to 99.6% at x = 25%. However, the bond angles, namely $\theta_1$ (O1-Mn2-O4) and $\theta_2$ (O1-Mn2-O1), constantly deviate from ideal tetrahedral value (109.47°) with increasing $Cr^{3+}$ doping, suggesting an evolving angular distortion of $Mn2@O_4$ tetrahedron. Interestingly, if defining $\Delta\theta = \theta_1 - \theta_2$ as the "distortion parameter", the plot of $\Delta\theta$ with respect to the Mn2-site vacancy number as shown in Figure 3c is surprisingly linear related. Therefore, this correspondence suggests that the increasing distortion of the $Mn2@O_4$ tetrahedron destabilizes $Mn^{3+}$ at this site, thereby promoting the Mn2-site vacancy formation as a means of relieving local lattice strain.

**Magnetic properties and doping-induced spin-glass state:** To investigate how the $Cr^{3+}$ site-preferential doping affects the magnetic properties of $KSb_3(Mn_{1-x}Cr_x)_9O_{19}$, temperature-dependent magnetic susceptibility ($\chi$) and its inverse ($\chi^{-1}$) are measured under an external magnetic field ($\mu_0H$) of 0.3 T and a zero-field cooling protocol. The diamagnetic background of the sample holder was collected for each sample and subtracted for better comparison. As can be seen in Figure 4a, $\chi$(T) and $\chi^{-1}$(T) of all samples exhibit similar behavior while they are offset for clarity. Typical Curie-Weiss (CW) behavior can be observed above ~ 100 K, suggesting paramagnetic state. A spike at ~ 45 K is found in $\chi$(T) up to $x$ = 25%, which was attributed to static magnetic moment or partial spin freezing in undoped $KSb_3Mn_9O_{19}$.[40] Furthermore, two broad peaks exist at lower temperatures at ~ 40 K and ~ 11 K, which was claimed as a result of short-range magnetic ordering in $KSb_3Mn_9O_{19}$, evidenced by its strong magnetic diffuse scattering observed in neutron powder diffraction,[40] and is likely the key contributor in the Cr-doped samples.

The high-temperature CW behavior was fitted from 100 K to 300 K using the CW law

$$\chi = \frac{C}{T - \theta_{CW}}$$

where C is a temperature-independent constant and is related to the effective moment ($\mu_{eff}$) *via* $\mu_{eff} = \sqrt{8C}$, and $\theta_{CW}$ is the CW temperature. The fitted $\mu_{eff}$ per magnetic ion and $\theta_{CW}$ are summarized in Figure 4b. With $Cr^{3+}$ ($S$ = 3/2) doped on $Mn^{3+}$ ($S$ = 2) site, the $\mu_{eff}$ drops from ~5.4 $\mu_B$ to ~ 5.2 $\mu_B$, consistent with the smaller spin state of $Cr^{3+}$. Moreover, the single valence state of Mn and Sb ions are proved by XPS measurements shown in Figure S2, which further supports the existence of $Mn^{3+}$ instead of other possibilities. $\theta_{CW}$ are fitted to range from ~ -160 K to ~ -180 K, suggesting strong antiferromagnetic interaction. However, it shows a puzzling trend with a spike at $x$ = 6.67% and a broad hump starting at $x$ = 13.3%. It was known that Mn2 puckered honeycomb and Mn3 Kagome sublattices are dominant in hosting antiferromagnetic interactions in undoped $KSb_3Mn_9O_{19}$.[40] The crucial $Mn^{3+}$-$Mn^{3+}$ antiferromagnetic interactions were determined to be, in decreasing order of importance, Mn3-Mn3, Mn2-Mn3 and Mn2-Mn2. Therefore, the interatomic distances for Mn3-Mn3 ($d_{Mn3-Mn3}$) and Mn2-Mn3 ($d_{Mn2-Mn3}$) pairs are shown in Figure 4c, because this distance is broadly considered the major factor controlling the strength of magnetic exchange interactions. Here, $d_{Mn2-Mn3}$ shows a smooth decreasing trend with increasing $x$, consistent with the smaller $Cr^{3+}$ on Mn3 site and the monotonically decreasing occupancy on the Mn2 site. However, $d_{Mn3-Mn3}$ exhibits a sudden drop between $x$ = 13.3% and $x$ = 16.7%, which is expected to strengthen the antiferromagnetic coupling between $Mn^{3+}$ on the Mn3 site, consistent with the decreasing trend of $\theta_{CW}$ in the same samples. The sudden jump of $\theta_{CW}$ from $x$ = 10% to 13.3% may be attributed to the complex structural distortions with $Cr^{3+}$ doping, which tilts the spin state of $Mn^{3+}/Cr^{3+}$ such that the superexchange pathway is disrupted and leads to weaker antiferromagnetic coupling.

Furthermore, one of the major superexchange pathways for the antiferromagnetic interactions between Mn2 ions is Mn2-O1-Sb2-O1-Mn2, which is shown in Figure S3a. Given the fact that $Mn^{3+}$ exists on the Sb2 site, as discussed in Figure 2b, it is likely that the substitution with $Mn^{3+}$ will interfere with the superexchange interactions in a way proposed in Figure S3b, demonstrating a less indirect magnetic exchange pathway without virtual hopping through the $Sb^{3+}$ ions. Therefore, the more $Mn^{3+}$ occupying Sb2 site, the stronger the antiferromagnetic interaction. Given the lowest Mn on Sb2 content at $x$ = 6.67% in Figure 2b, it is plausible to believe that this can contribute to the sudden jump of the fitted $\theta_{CW}$ at the same $x$.

Field-dependent magnetization of $KSb_3(Mn_{1-x}Cr_x)_9O_{19}$ under 2 K is shown in Figure 4d where similar behaviors are observed across the series. No evidence for ferromagnetic ordering can be seen. The observation of ferromagnetic components due to the coercivity is consistent with the undoped $KSb_3Mn_9O_{19}$, which is attributed to static moments or short-range magnetic ordering.[40] The possibility of (canted) antiferromagnetic and ferrimagnetic ordering is excluded by the heat capacity measurements shown below.

To reveal the nature of the low-temperature magnetism, AC magnetic susceptibility ($\chi_{AC}$) measurements were conducted from $f_{AC}$ = 100 Hz to 9984 Hz under 5 – 50 K with both AC and DC magnetic fields of 10 Oe on $x$ = 10%, 16.7% and 25%, and $\chi_{AC}$'(T) is shown in Figure 4e and Figure S4. No frequency-dependent behavior can be observed for $x$ = 10% and 16.7%, consistent with undoped $KSb_3Mn_9O_{19}$.[40] However, an AC-frequency-driven peak shift towards high-temperature region for the peak at ~11 K was seen for $x$ = 25%, suggesting the emergence of a doping-induced spin-glass state, as observed in other spin-glass systems.[69–71]

**Heat capacity:** Heat capacity ($C_p$) measurements were conducted to interpret the evolution of magnetic properties in $KSb_3(Mn_{1-x}Cr_x)_9O_{19}$. Figure 5a presents $C_p$(T) under zero magnetic field where each curve is offset for clarity. No λ-anomaly can be seen in any samples, suggesting the absence of long-range magnetic ordering, consistent with magnetic properties measurements and undoped $KSb_3Mn_9O_{19}$.[40] The heat capacity of a crystalline sample can be represented as $C_p = C_{el} + C_{ph} + C_{mag}$, where $C_{el}$ is electronic contribution that is typically zero for insulating materials, i.e., this system; $C_{ph}$ is the phononic contribution and $C_{mag}$ is the magnetic contribution. To better understand the magnetic fluctuations/short-range ordering in this system, a Debye model is employed to describe the phonon background by fitting $C_p$(T) from 70 K to 100 K using the relation

$$C_p = 9nR[(\frac{T}{\theta_D})^3 \int_0^{\theta_{D1}/T} \frac{x^4 e^x}{(e^x - 1)^2} dx]$$

where n is the number of atoms per formula unit, R is gas constant, $\theta_D$ is the Debye temperature. It can be seen in Figure 5a that the high-temperature data can be well described by the Debye model, while deviations from the model emerge below 70 K and become prominent below ~50 K, consistent with the spike in $\chi$(T) at ~45 K, indicating enhanced magnetic fluctuations/short-range ordering. After subtraction of $C_{ph}$, $C_{mag}$ can be isolated and analyzed. As shown in Figure 5b, a broad peak can be observed centered around 11 K in $C_{mag}$/T(T), revealing released magnetic entropy when heating the system, consistent with the transition at ~11 K found in $\chi$(T). Additionally, magnetic entropy ($\Delta S_{mag}$) can be obtained using $\Delta S_{mag} = \int \Delta C_{mag}/T dT$, which is also plotted in Figure 5b. For a magnetic ion with a spin number of $S$ where spin-orbit coupling is minor, the expected $\Delta S_{mag}$ in a long-range magnetically ordered system is Rln(2$S$+1) where R is gas constant, e.g., 11.5 J/mol/K for $Cr^{3+}$ ($S$ = 3/2) and 13.4 J/mol/K for $Mn^{3+}$ ($S$ = 2). When normalizing the obtained $\Delta S_{mag}$/f.u. to each magnetic ion, as shown in Figure 5c, it varies from ~3.4 J/mol/K to ~6.9 J/mol/K, with a minimum at $x$ = 6.67% and maximum at $x$ = 16.7%, revealing potentially weakest magnetic fluctuations for $KSb_3(Mn_{0.9333}Cr_{0.0667})_9O_{19}$ and strongest in $KSb_3(Mn_{0.833}Cr_{0.167})_9O_{19}$. However, none of the $\Delta S_{mag}/mol_{(Mn+Cr)}$ reaches the expected long-range ordered values, confirming the absence of long-range magnetic ordering. Moreover, the possible reason why $\Delta S_{mag}/mol_{(Mn+Cr)}$ drops dramatically for $x$ = 25% may be attributed to the emergence of spin-glass state for $KSb_3(Mn_{0.75}Cr_{0.25})_9O_{19}$, which significantly weakens the magnetic fluctuations, and thus, leads to smaller magnetic entropy.

## *Conclusion*

Here in this paper, we performed systematic studies on Cr-doped $KSb_3Mn_9O_{19}$, one of the first Mn-based M-type hexaferrites systems with quasi-two-dimensional magnetic frustration reported. Through a combination of powder X-ray diffraction, single-crystal X-ray diffraction, SEM-EDS and STEM-EDS, we were able to conduct in-depth structural analyses. Interestingly, $Cr^{3+}$ was found to possess a strong site preference for the Mn3 Kagome site, while an evolving site vacancy and $Mn^{3+}/Sb^{3+}$ site mixing were observed in the surrounding sites. Possible mechanisms for these structural observations were proposed and crystal-field effect, chemical bonding and local strain were assigned as the most crucial factors controlling the cation distribution in this system. Magnetic properties and heat capacity were measured and discussed, while no major changes were found compared to the parent compound, except for a doping-induced spin-glass state in the heaviest doped sample. This work provides mechanistic evidence for the cation distribution in a widely studied oxide family with geometrically frustrated sublattices. Meanwhile, the knowledge obtained from this paper can provide further guidance on for achieving cation-ordered M-type hexaferrites, for better magnetic properties, including permanent magnets and quantum frustrated magnetism.

## *Acknowledgements*

D.C., E.W. and X.G. thank the startup funding from the University of Pittsburgh. Work performed in the University of Pittsburgh NanoScale Fabrication and Characterization Facility (RRID:SCR_05124) and services and instruments used in this project were graciously supported, in part, by the University of Pittsburgh.

## *Appendix A. Supplementary data*

Supplementary data to this article can be found online at xxxxxxx.

**Figure 1. a.** Powder XRD patterns for $KSb_3(Mn_{1-x}Cr_x)_9O_{19}$ (0% ≤ x ≤ 25%) offset for clarity. **b.** *a* and *c* lattice parameters from PXRD and SCXRD for the doped series. **c.** Crystal structure for the parent compound with the different Sb sites highlighted (left), crystal structure showing the three different Mn sublattices and their coordination environments (right).

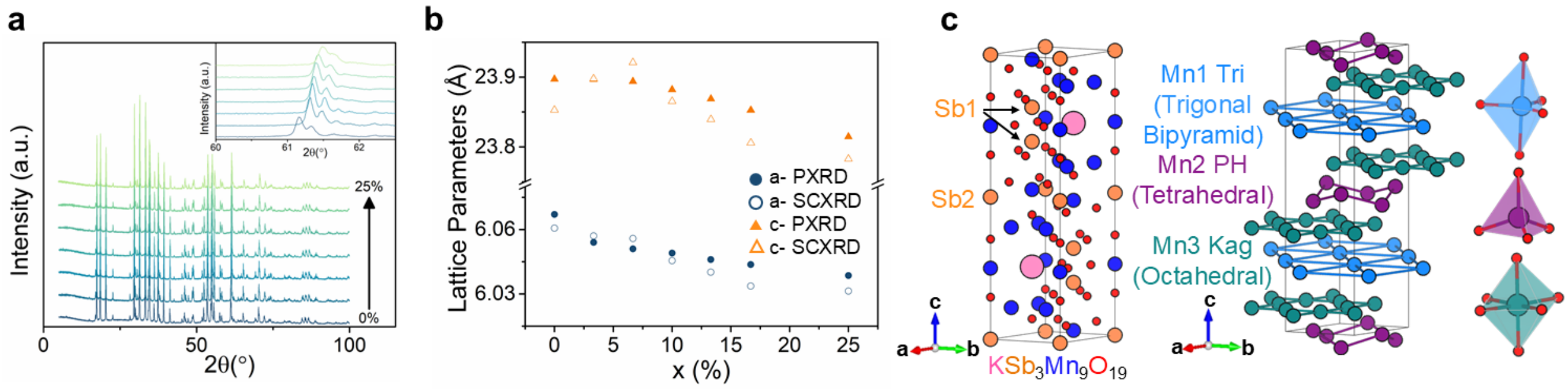

**Figure 2. a.** HAADF-STEM line profile analysis of the EDS maps of two different areas from x = 0.10. Teal Kag lines and purple PH lines along with the crystal structure added to help identify specific peaks. Element line colors: Blue = Mn, Green = Cr, Orange = Sb, Pink = K. **b.** % occupancy on the Mn2 site (bottom) and % of Mn on Sb2 site (top). **c.** Crystal structure showing the changes happening as Cr dopes into the system with the site-specific formula indicated. **d.** HAADF-STEM image of x = 0.10 for areas 1 and 2 (left), and HAADF-STEM and EDS mapping results used for line profile analysis (LPA) for area 1 LPA (right) with red arrow showing the direction of the LPA and crystal structure overlaid on select EDS maps for clarity.

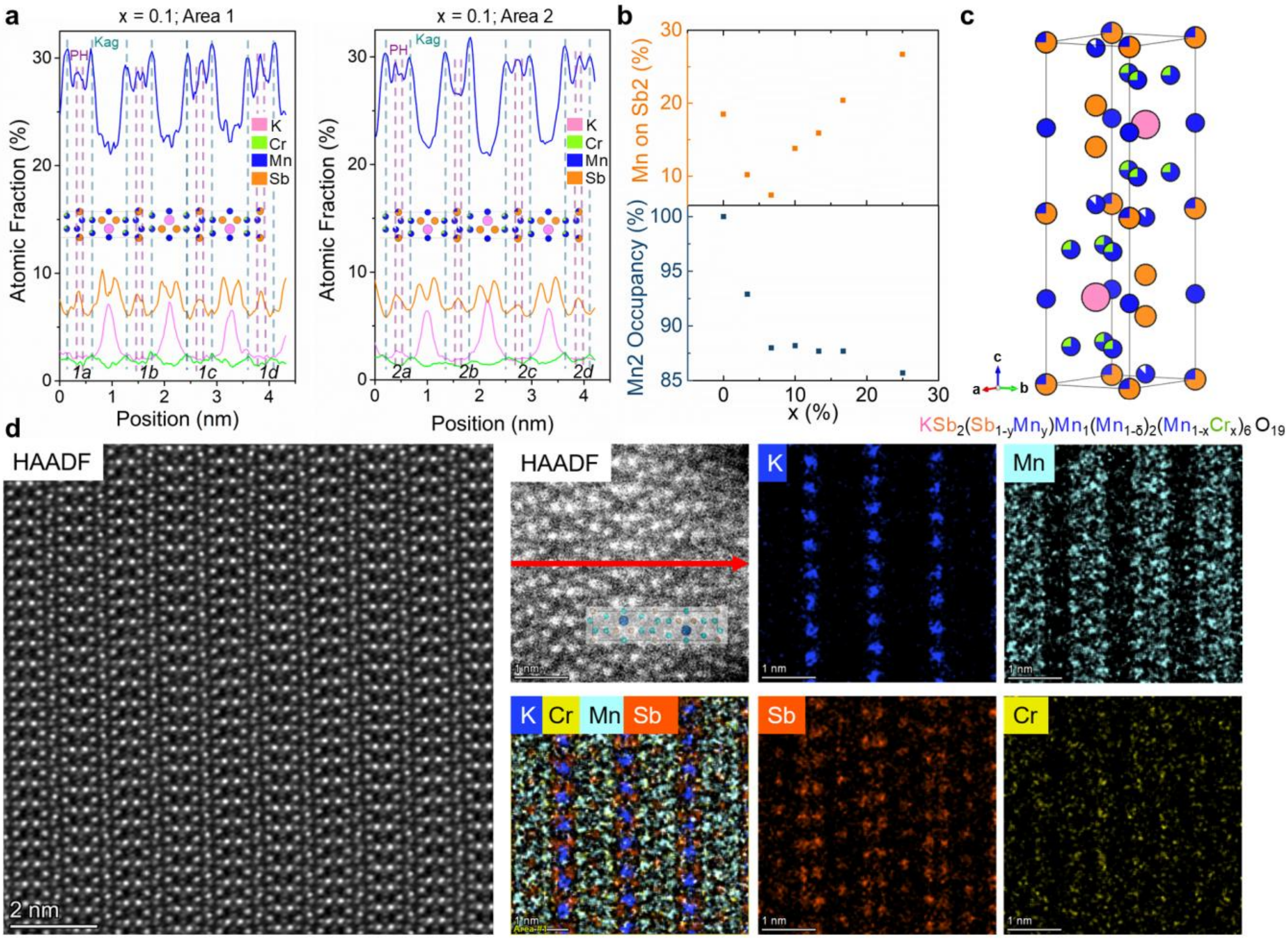

**Figure 3. a.** Electronic structure showing Cr occupying the three different Mn sites with calculated crystal field stabilization energy (CFSE) values. **b.** Crystal structure showing the region where bond lengths and bond angles were collected from where the different oxygen sites were color-coded. **c.** Cartoon showing the change in the $Mn2@O_4$ tetrahedron bond angles from the ideal to x =25% (left) and graph of the angle distortion ($\Delta\theta$) vs the vacancy on the Mn2 site with a line of best fit and $R^2$ value shown. Here, $\theta_1$ is the bond angle of O1-Mn2-O4, $\theta_2$ is the bond angle of O1-Mn2-O1, while $\Delta\theta = \theta1 - \theta2$.

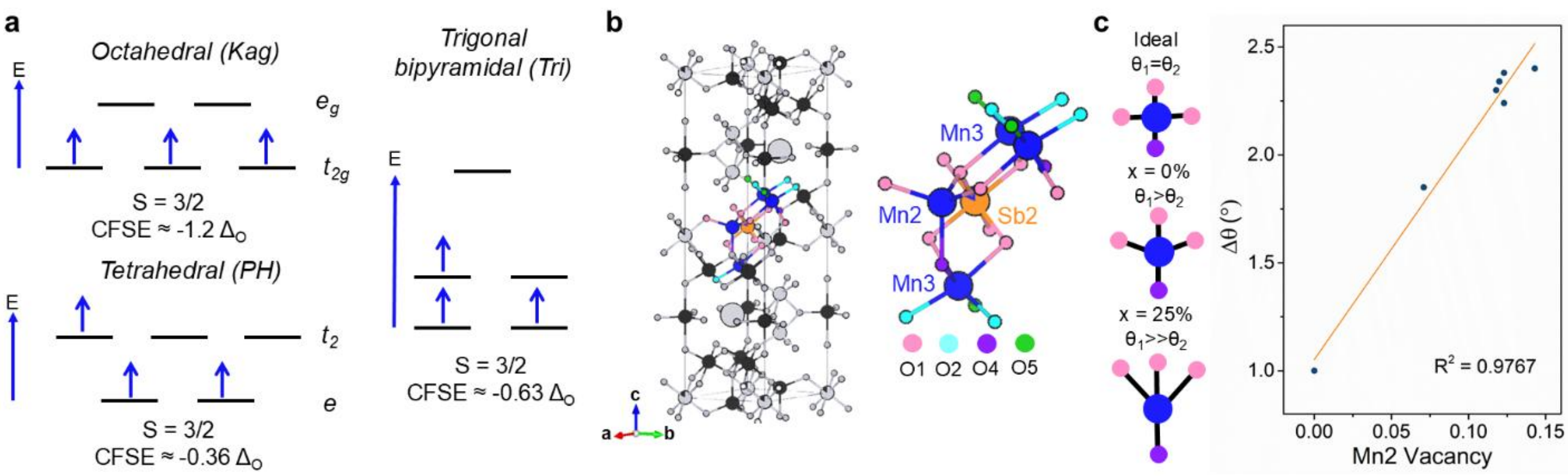

**Figure 4. a.** Magnetic susceptibility and its inverse measured in zero-field cooling protocol at 0.3 T for $KSb_3(Mn_{1-x}Cr_x)_9O_{19}$ ($0 \leq x \leq 0.25$), offset for clarity. **b.** The $\theta_{CW}$ (bottom) and $\mu_{eff}$ (top) for the doped series after fitting the 200 to 300 K region. **c.** Mn3-Mn3 bond lengths for the series (bottom) and Mn2-Mn3 bond lengths for the series (top). **d.** Hysteresis loops measured from -9 to 9 T for $KSb_3(Mn_{1-x}Cr_x)_9O_{19}$, the dashed lines represent offset for clarity. **e.** AC magnetic susceptibility for the x =25% Cr sample under various frequencies with curved arrow added to help show peak shift.

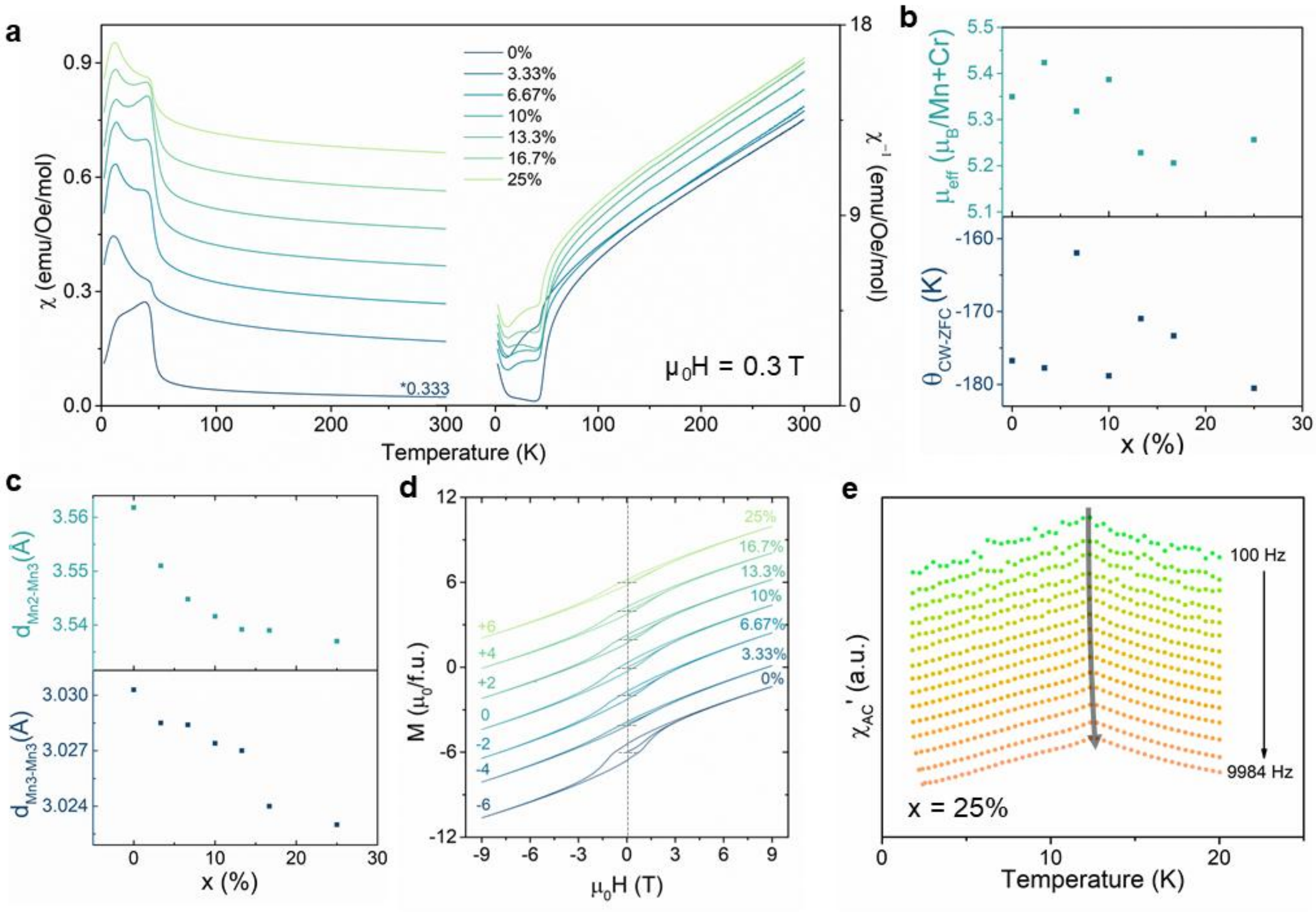

**Figure 5. a.** Zero-field heat capacity for $KSb_3(Mn_{1-x}Cr_x)_9O_{19}$ (0% ≤ x ≤ 25%), offset for clarity with Debye fitting shown as the orange line. **b.** The magnetic contribution to the heat capacity (left axis) for the doped series and the magnetic entropy (right axis) from 0% to 25% Cr . **c.** Saturated magnetic entropy per magnetic ion from 0% to 25% Cr with line added to help visualize trends.

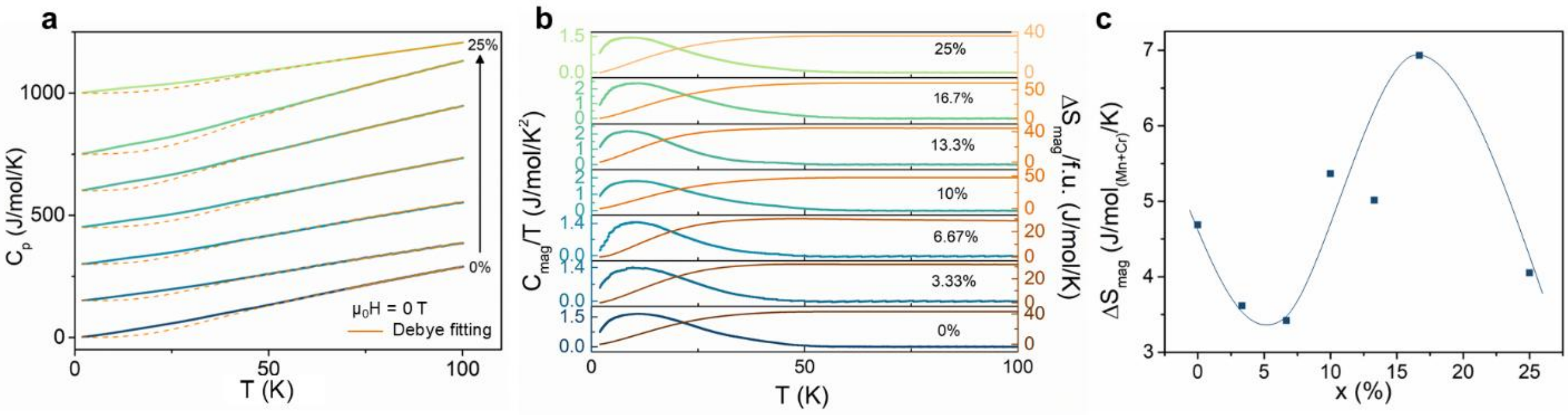

**Table 1.** Important metal-oxygen bond lengths for the full series of Cr-doped $KSb_3Mn_9O_{19}$.

| Cr Content (%) | Sb2-O1 | Mn3-O1 | Mn3-O2 | Mn3-O4 | Mn3-O5 | Mn2-O1 | Mn2-O4 |
|---|---|---|---|---|---|---|---|
| **0** | 1.977 | 2.163 | 2.04 | 2.013 | 1.984 | 2.061 | 2.106 |
| **3.33** | 1.974 | 2.169 | 2.041 | 2.008 | 1.982 | 2.066 | 2.104 |
| **6.67** | 1.974 | 2.166 | 2.041 | 2.001 | 1.986 | 2.073 | 2.115 |
| **10** | 1.974 | 2.154 | 2.039 | 2.005 | 1.988 | 2.068 | 2.098 |
| **13.3** | 1.975 | 2.144 | 2.036 | 1.995 | 1.984 | 2.068 | 2.116 |
| **16.7** | 1.973 | 2.134 | 2.034 | 1.999 | 1.983 | 2.07 | 2.105 |
| **25** | 1.973 | 2.123 | 2.035 | 2.008 | 1.989 | 2.076 | 2.084 |

**Table 2.** Changes on the distorted $Mn2@O_4$ tetrahedron with $Cr^{3+}$ doping

| Cr (%) | Mn2 (%) | Vacancy (%) | $\theta_1$ (O1-Mn2-O4) (°) | $\theta_2$ (O1-Mn2-O1) (°) | $\Delta\theta$ ($\theta_1$- $\theta_2$) (°) |
|---|---|---|---|---|---|
| **0** | 100 | 0 | 109.97 | 108.97 | 1 |
| **3.33** | 92.9 | 7.1 | 110.39 | 108.54 | 1.85 |
| **6.67** | 88 | 12 | 110.63 | 108.29 | 2.34 |
| **10** | 88.2 | 11.8 | 110.61 | 108.31 | 2.3 |
| **13.3** | 87.7 | 12.3 | 110.65 | 108.27 | 2.38 |
| **16.7** | 87.7 | 12.3 | 110.58 | 108.34 | 2.24 |
| **25** | 85.7 | 14.3 | 110.7 | 108.3 | 2.4 |

*Supplementary Information*

# Pronounced Site Preference in Cr-Doped Mn-Based M-Type Hexaferrites and Its Chemical Origins

*Dylan Correll,[a] Susheng Tan,[b] Evan Wang[a] and Xin Gui[a]**

[a] Department of Chemistry, University of Pittsburgh, Pittsburgh, PA, 15260, USA
[b] Department of Electrical and Computer Engineering and Gertrude E. and John M. Petersen Institute of Nanoscience and Engineering, University of Pittsburgh, Pittsburgh, PA, 15260, USA

**Table of Contents**

**Table S1.** Single crystal structure refinement for $KSb_3(Mn_{1-x}Cr_x)_9O_{19}$.

| Nominal Composition | x = 3.33% | x = 6.67% | x = 10% |
|---|---|---|---|
| Temperature (K) | 304(2) | 301(2) | 302(2) |
| F.W. (g/mol) | 1185.97 | 1185.44 | 1176.21 |
| Space group; Z | $P6_3/mmc$; 2 | $P6_3/mmc$; 2 | $P6_3/mmc$; 2 |
| $a$(Å) | 6.0570 (2) | 6.0558 (4) | 6.04550(14) |
| $c$(Å) | 23.8978 (2) | 23.921 (3) | 23.8653(8) |
| V (Å$^3$) | 759.28 (7) | 759.71(14) | 755.37(4) |
| θ range (°) | 3.410-33.147 | 3.407-33.741 | 3.415-33.148 |
| No. reflections; $R_{int}$ | 16279; 0.0787 | 18152; 0.0973 | 16947; 0.1027 |
| No. independent reflections | 618 | 648 | 617 |
| No. parameters | 34 | 34 | 34 |
| $R_1$: $\omega R_2$ ($I$>2δ($I$)) | 0.0247:0.0477 | 0.0240:0.0474 | 0.0304:0.0694 |
| Goodness of fit | 1.046 | 1.065 | 1.121 |
| Diffraction peak and hole (e$^-$/ Å$^3$) | 1.165; -1.207 | 0.969; -1.261 | 1.019; -1.096 |

**Table S1 (cont.).** Single crystal structure refinement for $KSb_3(Mn_{1-x}Cr_x)_9O_{19}$.

| Nominal Composition | x = 13.3% | x = 16.7% | x = 25% |
|---|---|---|---|
| Temperature (K) | 304(2) | 302(2) | 302(2) |
| F.W. (g/mol) | 1173.17 | 1167.41 | 1158.73 |
| Space group; Z | *P6₃/mmc;* 2 | *P6₃/mmc;* 2 | *P6₃/mmc;* 2 |
| *a*(Å) | 6.04010(14) | 6.03360(14) | 6.03130(14) |
| *c*(Å) | 23.8395(8) | 23.8057(10) | 23.7829(11) |
| V (Å$^3$) | 753.21(4) | 750.52(5) | 749.23(5) |
| θ range (°) | 3.418-33.144 | 3.423-33.163 | 3.427-33.125 |
| No. reflections; $R_{int}$ | 16191; 0.1246 | 16183; 0.1170 | 13503; 0.1189 |
| No. independent reflections | 616 | 612 | 609 |
| No. parameters | 34 | 34 | 34 |
| $R_1$: $\omega R_2$ ($I$>2δ($I$)) | 0.0293:0.512 | 0.0392:0.0772 | 0.0551:0.0926 |
| Goodness of fit | 1.065 | 1.117 | 1.213 |
| Diffraction peak and hole (e$^-$/ Å$^3$) | 1.059; -1.046 | 1.096; -1.234 | 1.391; -1.370 |

**Table S2.** Atomic coordinates and equivalent isotropic displacement parameters for $KSb_3(Mn_{1-x}Cr_x)_9O_{19}$. ($U_{eq}$ is defined as one-third of the trace of the orthogonalized $U_{ij}$ tensor (Å$^2$))

**x = 3.33%**

| Atom | Wyck. | Occ. | *x* | *y* | *z* | $U_{eq}$ |
|---|---|---|---|---|---|---|
| Sb1 | *4f* | 1 | 0.3333 | 0.6667 | 0.1897(1) | 0.0074(1) |
| Sb2 | *2a* | 0.898(7) | 1 | 0 | 0 | 0.0068(2) |
| Mn4 | *2a* | 0.102(7) | 1 | 0 | 0 | 0.0068(2) |
| Mn2 | *4f* | 0.929(6) | 0.3333 | 0.6667 | 0.0205(1) | 0.0119(3) |
| Mn3 | *12k* | 0.992(3) | 0.8334(1) | 0.1666(1) | 0.1089(1) | 0.0088(2) |
| Mn1 | *2b* | 1 | 1 | 0 | 0.25 | 0.0204(4) |
| K1 | *2d* | 1 | 0.6667 | 0.3333 | 0.25 | 0.0162(4) |
| O1 | *12k* | 1 | 1.1487(3) | 0.8513(3) | 0.0506(1) | 0.0183(7) |
| O2 | *12k* | 1 | 0.4950(4) | 0.5050(4) | 0.1510(1) | 0.0138(6) |
| O3 | *6h* | 1 | 0.1941(4) | 0.3882(8) | 0.25 | 0.0106(8) |
| O4 | *4f* | 1 | 0.6667 | 0.3333 | 0.0676(2) | 0.0193(12) |
| O5 | *4e* | 1 | 1 | 0 | 0.1479(3) | 0.0218(13) |

**x = 6.67%**

| Atom | Wyck. | Occ. | *x* | *y* | *z* | $U_{eq}$ |
|---|---|---|---|---|---|---|
| Sb1 | *4f* | 1 | 0.3333 | 0.6667 | 0.1897(1) | 0.0069(1) |
| Sb2 | *2a* | 0.927(7) | 1 | 0 | 0 | 0.0064(2) |
| Mn4 | *2a* | 0.073(7) | 1 | 0 | 0 | 0.0064(2) |
| Mn2 | *4f* | 0.904(6) | 0.3333 | 0.6667 | 0.0202(1) | 0.0112(3) |
| Mn3 | *12k* | 0.995(3) | 0.8334(2) | 0.1666(2) | 0.1089(2) | 0.0083(2) |
| Mn1 | *2b* | 1 | 1 | 0 | 0.25 | 0.0190(4) |
| K1 | *2d* | 1 | 0.6667 | 0.3333 | 0.25 | 0.0158(4) |
| O1 | *12k* | 1 | 1.1484(4) | 0.8516(4) | 0.0508(1) | 0.0184(7) |
| O2 | *12k* | 1 | 0.4949(4) | 0.5051(4) | 0.1510(1) | 0.0132(6) |
| O3 | *6h* | 1 | 0.1937(4) | 0.3874(8) | 0.25 | 0.0112(8) |
| O4 | *4f* | 1 | 0.6667 | 0.3333 | 0.0682(2) | 0.0172(11) |
| O5 | *4e* | 1 | 1 | 0 | 0.1483(2) | 0.0193(12) |

**Table S2 (cont.).** Atomic coordinates and equivalent isotropic displacement parameters for $KSb_3(Mn_{1-x}Cr_x)_9O_{19}$. ($U_{eq}$ is defined as one-third of the trace of the orthogonalized $U_{ij}$ tensor (Å$^2$))

**x = 10%**

| Atom | Wyck. | Occ. | *x* | *y* | *z* | $U_{eq}$ |
|---|---|---|---|---|---|---|
| Sb1 | *4f* | 1 | 0.3333 | 0.6667 | 0.1897(2) | 0.0088(2) |
| Sb2 | *2a* | 0.863(9) | 1 | 0 | 0 | 0.0080(2) |
| Mn4 | *2a* | 0.137(9) | 1 | 0 | 0 | 0.0080(2) |
| Mn2 | *4f* | 0.882(7) | 0.3333 | 0.6667 | 0.0204(2) | 0.0124(5) |
| Mn3 | *12k* | 0.985(4) | 0.8336(8) | 0.1664(8) | 0.1086(3) | 0.0096(2) |
| Mn1 | *2b* | 1 | 1 | 0 | 0.25 | 0.0210(5) |
| K1 | *2d* | 1 | 0.6667 | 0.3333 | 0.25 | 0.0174(6) |
| O1 | *12k* | 1 | 1.1485(4) | 0.8515(4) | 0.0509(2) | 0.0187(9) |
| O2 | *12k* | 1 | 0.4955(5) | 0.5045(5) | 0.1510(2) | 0.0154(8) |
| O3 | *6h* | 1 | 0.1930(5) | 0.3860(10) | 0.25 | 0.0111(10) |
| O4 | *4f* | 1 | 0.6667 | 0.3333 | 0.0675(3) | 0.0194(14) |
| O5 | *4e* | 1 | 1 | 0 | 0.1487(3) | 0.0226(16) |

**x = 13.3%**

| Atom | Wyck. | Occ. | *x* | *y* | *z* | $U_{eq}$ |
|---|---|---|---|---|---|---|
| Sb1 | *4f* | 1 | 0.3333 | 0.6667 | 0.1896(2) | 0.0080(2) |
| Sb2 | *2a* | 0.841(9) | 1 | 0 | 0 | 0.0079(3) |
| Mn4 | *2a* | 0.159(9) | 1 | 0 | 0 | 0.0079(3) |
| Mn2 | *4f* | 0.877(7) | 0.3333 | 0.6667 | 0.02065(7) | 0.0109(5) |
| Mn3 | *12k* | 0.984(4) | 0.8337(2) | 0.1663(2) | 0.1084(3) | 0.0088(2) |
| Mn1 | *2b* | 1 | 1 | 0 | 0.25 | 0.0194(5) |
| K1 | *2d* | 1 | 0.6667 | 0.3333 | 0.25 | 0.0159(6) |
| O1 | *12k* | 1 | 1.1484(5) | 0.8516(5) | 0.0512(2) | 0.0208(10) |
| O2 | *12k* | 1 | 0.4962(5) | 0.5038(5) | 0.1509(2) | 0.0147(8) |
| O3 | *6h* | 1 | 0.1941(5) | 0.3882(11) | 0.25 | 0.0105(11) |
| O4 | *4f* | 1 | 0.6667 | 0.3333 | 0.0681(3) | 0.0175(15) |
| O5 | *4e* | 1 | 1 | 0 | 0.1485(3) | 0.0185(15) |

**Table S2 (cont.).** Atomic coordinates and equivalent isotropic displacement parameters for $KSb_3(Mn_{1-x}Cr_x)_9O_{19}$. ($U_{eq}$ is defined as one-third of the trace of the orthogonalized $U_{ij}$ tensor (Å$^2$))

**x = 16.7%**

| **Atom** | Wyck. | Occ. | *x* | *y* | *z* | $U_{eq}$ |
|---|---|---|---|---|---|---|
| Sb1 | *4f* | 1 | 0.3333 | 0.6667 | 0.1895(3) | 0.0093(2) |
| Sb2 | *2a* | 0.796(11) | 1 | 0 | 0 | 0.0092(3) |
| Mn4 | *2a* | 0.204(11) | 1 | 0 | 0 | 0.0092(3) |
| Mn2 | *4f* | 0.877(9) | 0.3333 | 0.6667 | 0.0209(2) | 0.0127(6) |
| Mn3 | *12k* | 0.975(5) | 0.8337(1) | 0.1663(1) | 0.1084(4) | 0.0106(3) |
| Mn1 | *2b* | 1 | 1 | 0 | 0.25 | 0.0216(6) |
| K1 | *2d* | 1 | 0.6667 | 0.3333 | 0.25 | 0.0179(7) |
| O1 | *12k* | 1 | 1.1479(6) | 0.8521(6) | 0.0515(2) | 0.0207(11) |
| O2 | *12k* | 1 | 0.4960(6) | 0.5040(6) | 0.1508(2) | 0.0155(9) |
| O3 | *6h* | 1 | 0.1934(6) | 0.3868(13) | 0.25 | 0.0102(12) |
| O4 | *4f* | 1 | 0.6667 | 0.3333 | 0.0675(4) | 0.0201(18) |
| O5 | *4e* | 1 | 1 | 0 | 0.1485(4) | 0.022(2) |

**x = 25%**

| **Atom** | **Wyck.** | **Occ.** | ***x*** | ***y*** | ***z*** | $U_{eq}$ |
|---|---|---|---|---|---|---|
| Sb1 | *4f* | 1 | 0.3333 | 0.6667 | 0.1894(4) | 0.0100(3) |
| Sb2 | *2a* | 0.733(16) | 1 | 0 | 0 | 0.0092(5) |
| Mn4 | *2a* | 0.267(16) | 1 | 0 | 0 | 0.0092(5) |
| Mn2 | *4f* | 0.857(12) | 0.3333 | 0.6667 | 0.0211(1) | 0.0123(8) |
| Mn3 | *12k* | 0.968(7) | 0.8337(1) | 0.1662(2) | 0.1082(5) | 0.0107(3) |
| Mn1 | *2b* | 1 | 1 | 0 | 0.25 | 0.0224(9) |
| K1 | *2d* | 1 | 0.6667 | 0.3333 | 0.25 | 0.0186(10) |
| O1 | *12k* | 1 | 1.1474(8) | 0.8526(8) | 0.0519(3) | 0.0208(16) |
| O2 | *12k* | 1 | 0.4960(9) | 0.5040(9) | 0.1509(2) | 0.0156(12) |
| O3 | *6h* | 1 | 0.1941(9) | 0.3882(18) | 0.25 | 0.0110(17) |
| O4 | *4f* | 1 | 0.6667 | 0.3333 | 0.0665(5) | 0.022(3) |
| O5 | *4e* | 1 | 1 | 0 | 0.1490(6) | 0.024(3) |

**Table S3.** Anisotropic thermal displacement parameters for $KSb_3(Mn_{1-x}Cr_x)_9O_{19}$.

**x = 3.33%**

| Atom | U11 | U22 | U33 | U23 | U13 | U12 |
|---|---|---|---|---|---|---|
| Sb1 | 0.0083(2) | 0.0083(2) | 0.0057(2) | 0 | 0 | 0.0041(2) |
| Sb2 | 0.0075(2) | 0.0075(2) | 0.0053(3) | 0 | 0 | 0.0037(1) |
| Mn4 | 0.0075(2) | 0.0075(2) | 0.0053(3) | 0 | 0 | 0.0037(1) |
| Mn2 | 0.0122(4) | 0.0122(4) | 0.0113(6) | 0 | 0 | 0.0061(2) |
| Mn3 | 0.0095(2) | 0.0095(2) | 0.0094(3) | -0.0001(1) | 0.0001(2) | 0.0063(2) |
| Mn1 | 0.0084(4) | 0.0084(4) | 0.0443(12) | 0 | 0 | 0.0042(2) |
| K1 | 0.0162(6) | 0.0162(6) | 0.0162(10) | 0 | 0 | 0.0081(3) |

**x = 6.67%**

| Atom | U11 | U22 | U33 | U23 | U13 | U12 |
|---|---|---|---|---|---|---|
| Sb1 | 0.0076(2) | 0.0076(2) | 0.0054(2) | 0 | 0 | 0.0038(7) |
| Sb2 | 0.0071(2) | 0.0071(2) | 0.0049(3) | 0 | 0 | 0.0035(1) |
| Mn4 | 0.0071(2) | 0.0071(2) | 0.0049(3) | 0 | 0 | 0.0035(1) |
| Mn2 | 0.0112(4) | 0.0112(4) | 0.0112(6) | 0 | 0 | 0.0056(2) |
| Mn3 | 0.0088(2) | 0.0088(2) | 0.0090(3) | -0.0001(1) | 0.0001(1) | 0.0057(2) |
| Mn1 | 0.0070(4) | 0.0070(4) | 0.0431(11) | 0 | 0 | 0.0035(2) |
| K1 | 0.0155(6) | 0.0155(6) | 0.0165(10) | 0 | 0 | 0.0078(3) |

**x = 10%**

| Atom | U11 | U22 | U33 | U23 | U13 | U12 |
|---|---|---|---|---|---|---|
| Sb1 | 0.0089(2) | 0.0089(2) | 0.0062(2) | 0 | 0 | 0.0044(1) |
| Sb2 | 0.0086(3) | 0.0086(3) | 0.0063(4) | 0 | 0 | 0.0043(2) |
| Mn4 | 0.0086(3) | 0.0086(3) | 0.0063(4) | 0 | 0 | 0.0043(2) |
| Mn2 | 0.0109(6) | 0.0109(6) | 0.0109(8) | 0 | 0 | 0.0055(3) |
| Mn3 | 0.0092(3) | 0.0092(3) | 0.0095(4) | 0.0003(2) | -0.0002(1) | 0.0057(3) |
| Mn1 | 0.0072(6) | 0.0072(6) | 0.0438(15) | 0 | 0 | 0.0036(3) |
| K1 | 0.0152(8) | 0.0152(8) | 0.0174(13) | 0 | 0 | 0.0076(4) |

**Table S3 (cont.).** Anisotropic thermal displacement parameters for $KSb_3(Mn_{1-x}Cr_x)_9O_{19}$.

**x = 13.3%**

| Atom | U11 | U22 | U33 | U23 | U13 | U12 |
|---|---|---|---|---|---|---|
| Sb1 | 0.0086(2) | 0.0086(2) | 0.0091(2) | 0 | 0 | 0.0043(1) |
| Sb2 | 0.0078(3) | 0.0078(3) | 0.0084(4) | 0 | 0 | 0.0039(2) |
| Mn4 | 0.0078(3) | 0.0078(3) | 0.0084(4) | 0 | 0 | 0.0039(2) |
| Mn2 | 0.0123(6) | 0.0123(6) | 0.0126(7) | 0 | 0 | 0.0062(3) |
| Mn3 | 0.0091(3) | 0.0091(3) | 0.0125(4) | 0.0001(2) | -0.0001(1) | 0.0058(3) |
| Mn1 | 0.0078(5) | 0.0078(5) | 0.0475(15) | 0 | 0 | 0.0039(3) |
| K1 | 0.0165(8) | 0.0165(8) | 0.0191(12) | 0 | 0 | 0.0083(4) |

**x = 16.7%**

| Atom | U11 | U22 | U33 | U23 | U13 | U12 |
|---|---|---|---|---|---|---|
| Sb1 | 0.0090(2) | 0.0090(2) | 0.0099(3) | 0 | 0 | 0.0045(1) |
| Sb2 | 0.0089(4) | 0.0089(4) | 0.0098(5) | 0 | 0 | 0.0045(2) |
| Mn4 | 0.0089(4) | 0.0089(4) | 0.0098(5) | 0 | 0 | 0.0045(2) |
| Mn2 | 0.0125(7) | 0.0125(7) | 0.0132(9) | 0 | 0 | 0.0062(3) |
| Mn3 | 0.0097(4) | 0.0097(4) | 0.0134(4) | 0.0000(2) | 0.0000(2) | 0.0057(4) |
| Mn1 | 0.0082(7) | 0.0082(7) | 0.0483(19) | 0 | 0 | 0.0041(3) |
| K1 | 0.0164(10) | 0.0164(10) | 0.0208(16) | 0 | 0 | 0.0082(5) |

**x = 25%**

| Atom | U11 | U22 | U33 | U23 | U13 | U12 |
|---|---|---|---|---|---|---|
| Sb1 | 0.0103(3) | 0.0103(3) | 0.0095(4) | 0 | 0 | 0.0051(2) |
| Sb2 | 0.0091(6) | 0.0091(6) | 0.0095(7) | 0 | 0 | 0.0045(3) |
| Mn4 | 0.0091(6) | 0.0091(6) | 0.0095(7) | 0 | 0 | 0.0045(3) |
| Mn2 | 0.0132(10) | 0.0132(10) | 0.0104(12) | 0 | 0 | 0.0066(5) |
| Mn3 | 0.0098(5) | 0.0098(5) | 0.0135(6) | 0.0006(3) | -0.0006(3) | 0.0057(5) |
| Mn1 | 0.0092(10) | 0.0092(10) | 0.049(3) | 0 | 0 | 0.0046(5) |
| K1 | 0.0182(15) | 0.0182(15) | 0.020(2) | 0 | 0 | 0.0091(7) |

**Table S4.** Lower-than-unity occupancies for the Mn2, Mn3, and Sb2 sites.

| Cr Content (%) | Mn2 Site | Mn3 Site | Sb2 Site |
|---|---|---|---|
| 3.33 | 0.929(6) | 0.992(3) | 0.898(7) |
| 6.67 | 0.904(6) | 0.995(3) | 0.927(7) |
| 10 | 0.882(7) | 0.985(4) | 0.863(9) |
| 13.3 | 0.877(7) | 0.984(4) | 0.841(9) |
| 16.7 | 0.877(9) | 0.975(5) | 0.796(11) |
| 25 | 0.857(12) | 0.968(7) | 0.733(16) |

**Figure S1.** HAADF-STEM and EDS mapping results used for line profile analysis (LPA) for area 2 LPA (right) with red arrow showing the direction of the LPA and crystal structure overlaid on select EDS maps for clarity.

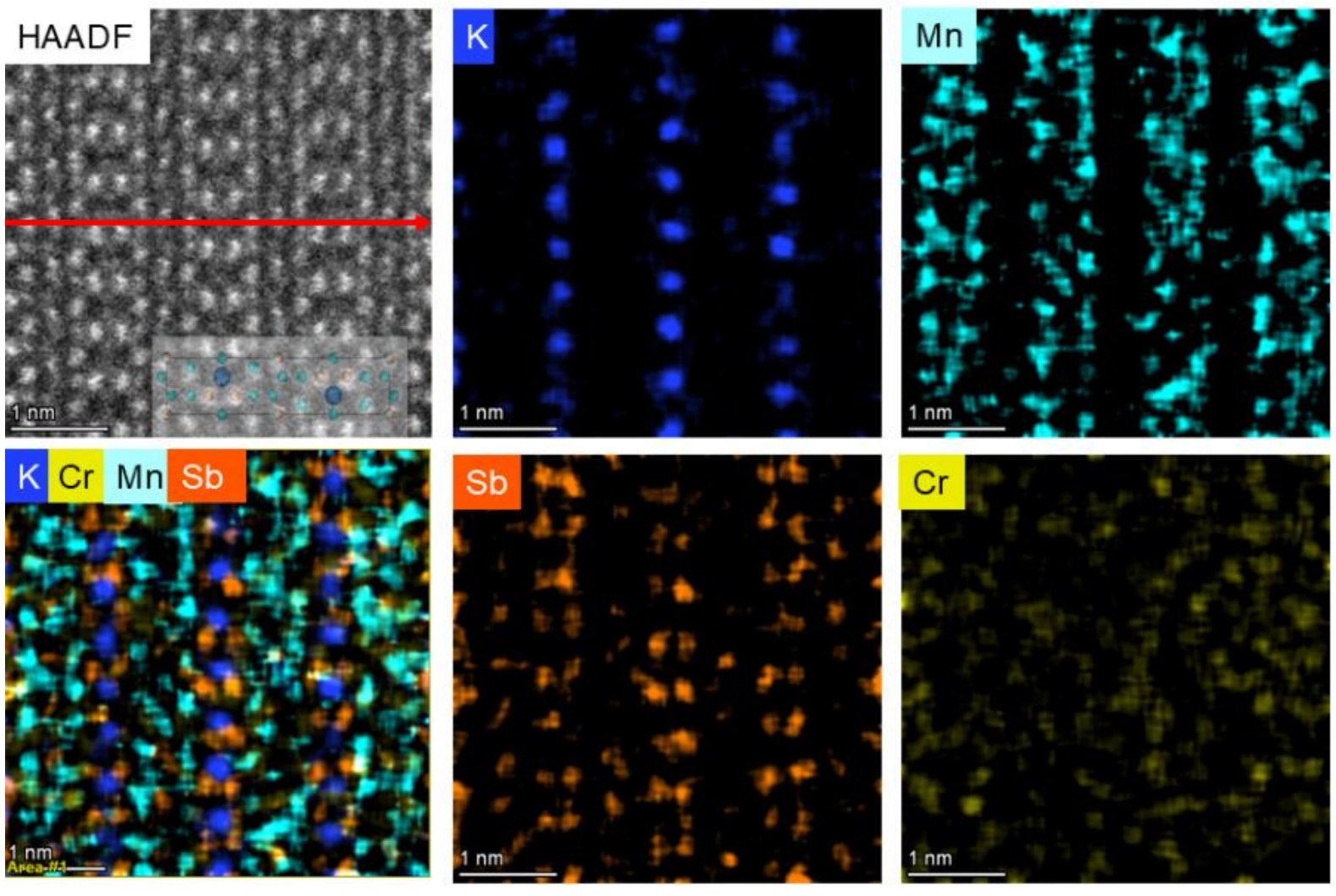

**Figure S2. a.** XPS spectra for Sb ions in $KSb_3Mn_9O_{19}$. **b.** Mn ions in $KSb_3Mn_9O_{19}$. **c.** Sb ions in $KSb_3(Mn_{0.75}Cr_{0.25})_9O_{19}$. **d.** Mn ions in $KSb_3(Mn_{0.75}Cr_{0.25})_9O_{19}$.

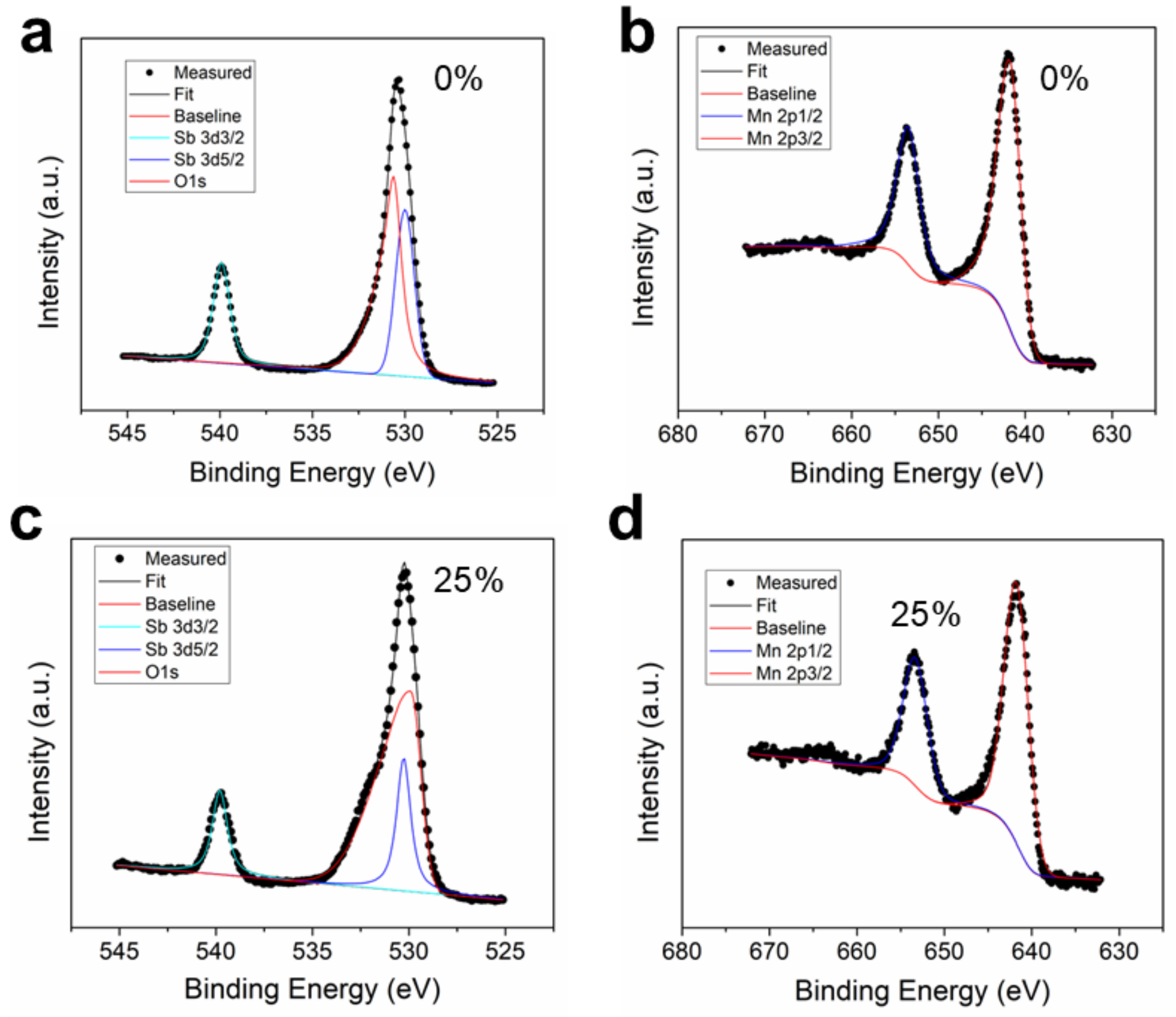

**Figure S3. a.** Proposed superexchange pathway for antiferromagnetic interactions between Mn2 ions. **b.** Proposed superexchange pathway when $Mn^{3+}$ occupies the Sb2 site. **c.** Structure showing superexchange pathway for Mn2 ions.

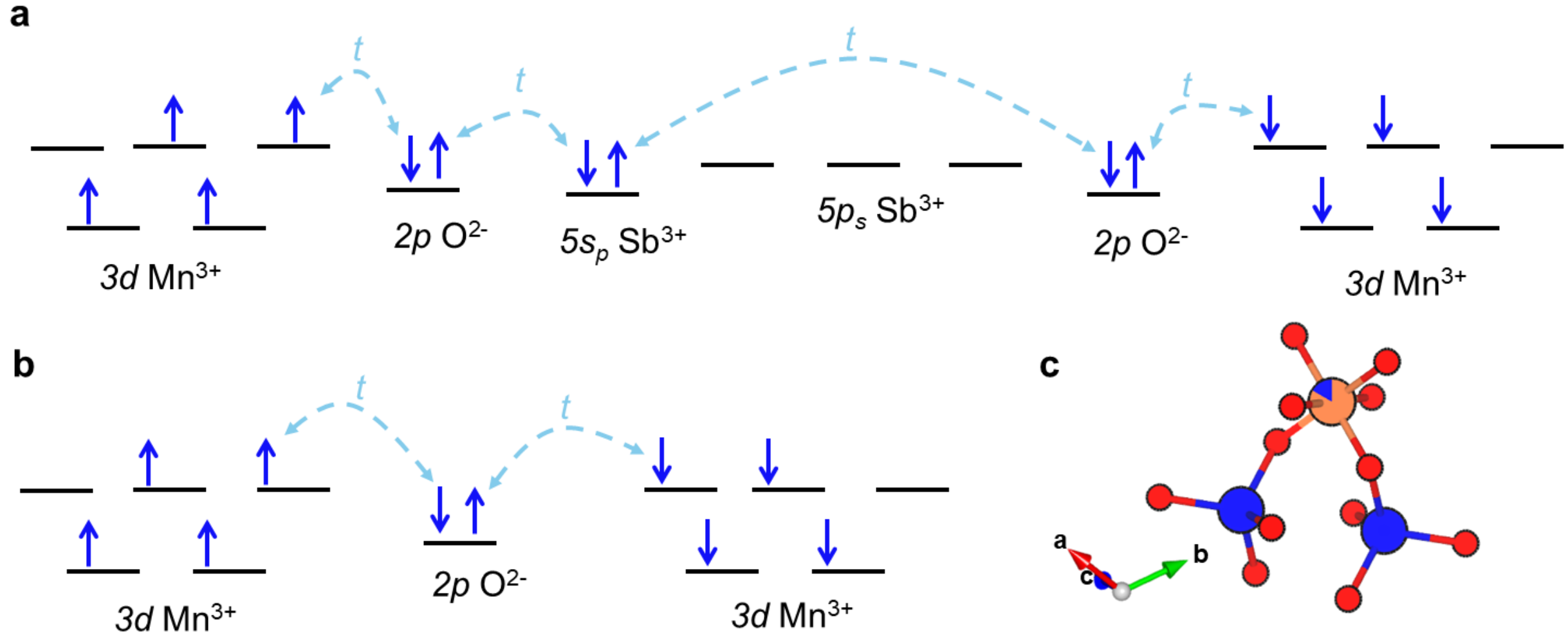

**Figure S4.** AC magnetic susceptibility for the x = 10% Cr, **a**, 16.7%, **b**, and 25%, **c**, samples under various frequencies.

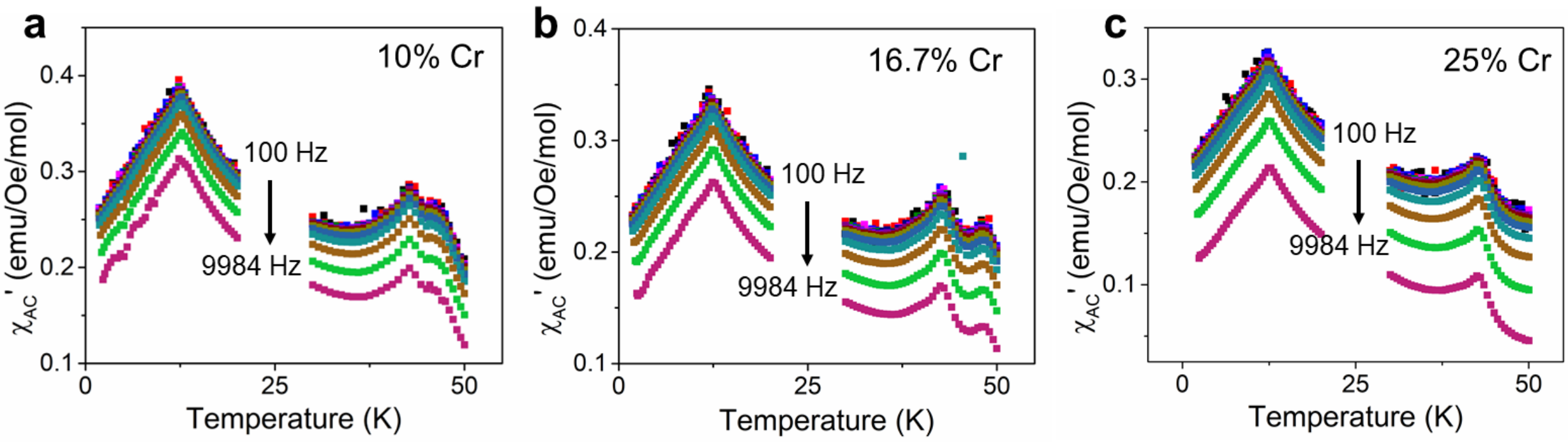